\documentclass[aps,prc,twocolumn,superscriptaddress,showpacs]{revtex4}
\usepackage{graphicx} 
\usepackage{multirow}

\begin{document}


\title{Coulomb excitation of ${}^{73}$Ga}


\author{J.~Diriken}
\affiliation{Instituut voor Kern- en Stralingsfysica, K.U.Leuven, Celestijnenlaan 200D, B-3001 Leuven, Belgium}
\author{I.~Stefanescu}
\affiliation{Instituut voor Kern- en Stralingsfysica, K.U.Leuven, Celestijnenlaan 200D, B-3001 Leuven, Belgium}
\affiliation{Department of Chemistry and Biochemistry, University of Maryland, College Park, Maryland 20742, USA}
\author{D.~Balabanski}
\affiliation{Dipartimento di Fisica, Universita di Camerino, I-62032 Camerino, Italy}
\affiliation{INRNE, Bulgarian Academy of Science, BG-1784 Sofia, Bulgaria}
\author{N.~Blasi}
\affiliation{Dipartimento di Fisica, Universita di Camerino, I-62032 Camerino, Italy}
\author{A.~Blazhev}
\affiliation{IKP, University of Cologne, D-50937 Cologne, Germany}
\author{N.~Bree}
\affiliation{Instituut voor Kern- en Stralingsfysica, K.U.Leuven, Celestijnenlaan 200D, B-3001 Leuven, Belgium}
\author{J.~Cederk\"{a}ll}
\affiliation{ISOLDE, CERN, CH-1211 Geneva 23, Switserland}
\affiliation{Physics Department, University of Lund, Box-118, SE-22100, Lund, Sweden}
\author{T.E.~Cocolios}
\affiliation{Instituut voor Kern- en Stralingsfysica, K.U.Leuven, Celestijnenlaan 200D, B-3001 Leuven, Belgium}
\author{T.~Davinson}
\affiliation{Department of Physics and Astronomy, University of Edinburgh, Edinburgh EH9 3JZ, United Kingdom}
\author{J.~Eberth}
\affiliation{IKP, University of Cologne, D-50937 Cologne, Germany}
\author{A.~Ekstr\"{o}m}
\affiliation{Physics Department, University of Lund, Box-118, SE-22100, Lund, Sweden}
\author{D.V.~Fedorov}
\affiliation{Petersburg Nuclear Physics Institute, 188300 Gatchina, Russia}
\author{V.N.~Fedosseev}
\affiliation{ISOLDE, CERN, CH-1211 Geneva 23, Switserland}
\author{L.M.~Fraile}
\affiliation{ISOLDE, CERN, CH-1211 Geneva 23, Switserland}
\affiliation{Grupo de F\'isica Nuclear, Universidad Complutense, E-28040 Madrid, Spain}
\author{S.~Franchoo}
\affiliation{IPN Orsay, F-91406 Orsay Cedex, France}
\author{G.~Georgiev}
\affiliation{CSNSM, CNRS/IN2P3; Universit\'e Paris-Sud, UM8609, F-91405 Orsay, France}
\author{K.~Gladnishki}
\affiliation{Faculty of Physics, University of Sofia, BG-1164, Sofia, Bulgaria}
\author{M.~Huyse}
\affiliation{Instituut voor Kern- en Stralingsfysica, K.U.Leuven, Celestijnenlaan 200D, B-3001 Leuven, Belgium}
\author{O.V.~Ivanov}
\affiliation{Instituut voor Kern- en Stralingsfysica, K.U.Leuven, Celestijnenlaan 200D, B-3001 Leuven, Belgium}
\author{V.S.~Ivanov}
\affiliation{Petersburg Nuclear Physics Institute, 188300 Gatchina, Russia}
\author{J.~Iwanicki}
\affiliation{Heavy Ion Laboratory, Warsaw University, Pasteura 5A, 02-093 Warsaw, Poland}
\author{J.~Jolie}
\affiliation{IKP, University of Cologne, D-50937 Cologne, Germany}
\author{T.~Konstantinopoulos}
\affiliation{National Research Center Demokritos, Greece}
\author{Th.~Kr\"{o}ll\footnote{Present address: Technische Universit\"{a}t Darmstadt , Institut f\"{u}r Kernphysik , Schlossgartenstr. 9 D-64289 Darmstadt, Germany}}
\affiliation{Technische Universit\"{a}t M\"{u}nchen, Physik Department, D-85748 Garching, Germany}
\author{R.~Kr\"{u}cken}
\affiliation{Technische Universit\"{a}t M\"{u}nchen, Physik Department, D-85748 Garching, Germany}
\author{U.~K\"{o}ster}
\affiliation{ISOLDE, CERN, CH-1211 Geneva 23, Switserland}
\affiliation{Intitut Laue Langevin, 6 rue Jules Horowitz, F-38042 Grenoble, France}
\author{A.~Lagoyannis}
\affiliation{National Research Center Demokritos, Greece}
\author{G.~Lo Bianco}
\affiliation{Dipartimento di Fisica, Universita di Camerino, I-62032 Camerino, Italy}
\author{P.~Maierbeck}
\affiliation{Technische Universit\"{a}t M\"{u}nchen, Physik Department, D-85748 Garching, Germany}
\author{B.A~Marsh}
\affiliation{ISOLDE, CERN, CH-1211 Geneva 23, Switserland}
\author{P.~Napiorkowski}
\affiliation{Heavy Ion Laboratory, Warsaw University, Pasteura 5A, 02-093 Warsaw, Poland}
\author{N.~Patronis}
\affiliation{Instituut voor Kern- en Stralingsfysica, K.U.Leuven, Celestijnenlaan 200D, B-3001 Leuven, Belgium}
\author{D.~Pauwels}
\affiliation{Instituut voor Kern- en Stralingsfysica, K.U.Leuven, Celestijnenlaan 200D, B-3001 Leuven, Belgium}
\author{P.~Reiter}
\affiliation{IKP, University of Cologne, D-50937 Cologne, Germany}
\author{M.~Seliverstov}
\affiliation{Petersburg Nuclear Physics Institute, 188300 Gatchina, Russia}
\author{G.~Sletten}
\affiliation{The Niels Bohr Institute, University of Copenhagen, 2100 Copenhagen, Denmark}
\author{J.~Van de Walle}
\affiliation{Instituut voor Kern- en Stralingsfysica, K.U.Leuven, Celestijnenlaan 200D, B-3001 Leuven, Belgium}
\affiliation{ISOLDE, CERN, CH-1211 Geneva 23, Switserland}
\affiliation{Kernfysisch Versneller Instituut, University of Groningen, NL-9747 AA Groningen, the Netherlands}
\author{P.~Van Duppen}
\affiliation{Instituut voor Kern- en Stralingsfysica, K.U.Leuven, Celestijnenlaan 200D, B-3001 Leuven, Belgium}
\author{D.~Voulot}
\affiliation{ISOLDE, CERN, CH-1211 Geneva 23, Switserland}
\author{W.B.~Walters}
\affiliation{Department of Chemistry and Biochemistry, University of Maryland, College Park, Maryland 20742, USA}
\author{N.~Warr}
\affiliation{IKP, University of Cologne, D-50937 Cologne, Germany}
\author{F.~Wenander}
\affiliation{ISOLDE, CERN, CH-1211 Geneva 23, Switserland}
\author{K.~Wrzosek}
\affiliation{Heavy Ion Laboratory, Warsaw University, Pasteura 5A, 02-093 Warsaw, Poland}

\date{\today}

\begin{abstract}
The $B(E2;I_i\rightarrow I_f)$ values for transitions in ${}^{71}_{31}$Ga${}_{40}$ and ${}^{73}_{31}$Ga${}_{42}$ were deduced from a Coulomb excitation experiment at the safe energy of 2.95 MeV/nucleon using post-accelerated beams of ${}^{71,73}$Ga at the REX-ISOLDE on-line isotope mass separator facility. The emitted $\gamma$ rays were detected by the MINIBALL $\gamma$-detector array and $B(E2;I_i\rightarrow I_f)$ values were obtained from the yields normalized to the known strength of the $2^+ \rightarrow 0^+$ transition in the ${}^{120}$Sn target. The comparison of these new results with the data of less neutron-rich gallium isotopes shows a shift of the E2 collectivity towards lower excitation energy when adding neutrons beyond $N=40$. This supports conclusions from previous studies of the gallium isotopes which indicated a structural change in this isotopical chain between $N=40$ and $N=42$. Combined with recent measurements from collinear laser spectroscopy showing a $1/2^-$ spin and parity for the ground state, the extracted results revealed evidence for a $1/2^-,3/2^-$ doublet near the ground state in ${}^{73}_{31}$Ga${}_{42}$ differing by at most 0.8 keV in energy.
\end{abstract}


\maketitle

\section{INTRODUCTION}
The structure of the neutron rich odd-A gallium isotopes ($Z=31$) has been studied in the past by means of $\beta$ decay~\cite{Erdal1972449,Runte1983163,Zoller1970177}, single and multi-particle transfer reactions~\cite{PhysRevC.18.86,Riccato1974461,PhysRevC.19.1276,PhysRevC.21.2293,PhysRevC.2.149,PhysRevC.9.409} and more recently by deep inelastic reactions~\cite{irina_deep_inelastic}. The two neutron transfer reaction data suggest a change in structure between ${}^{71}$Ga${}_{40}$ and ${}^{73}$Ga$_{42}$, which is most probably related to the structural difference observed between ${}^{72}$Ge${}_{40}$ and ${}^{74}$Ge${}_{42}$ ($Z=32$), where both the ${}^{72}$Ge${}_{40}$(t,p)${}^{74}$Ge$_{42}$ and ${}^{74}$Ge$_{42}$(p,t)${}^{72}$Ge${}_{40}$ reactions strongly populate the excited $0^+$ state, indicating a shape transition~\cite{Vergnes1978447}. The two neutron transfer reaction ${}^{71}$Ga${}_{40}$(t,p)${}^{73}$Ga$_{42}$ indicated indeed a similar shift of the $\Delta L = 0$ strength as it is almost equally divided over three 3/2${}^-$ states at energies of 0, 219 and 915 keV~\cite{PhysRevC.19.1276}. 
\par
Recently a collinear laser spectroscopy measurement confirmed the assumption of a 3/2${}^-$ ground state in all odd-A gallium isotopes, except for ${}^{73}$Ga$_{42}$ and ${}^{81}$Ga${}_{50}$ where the spin and parity of the ground states were found to be 1/2${}^-$ and 5/2${}^-$, respectively~\cite{cheal}. 
\par
A combination of the results from previous experiments opens the possibility of a 1/2${}^{-}$, 3/2${}^{-}$ ground state doublet in ${}^{73}$Ga. 
The existence of such a doublet is compatible with the available transfer reaction data as the population of the yrast $1/2^-$ state is weak compared to the $3/2^-$ ground state in similar experiments studying ${}^{69,71}$Ga~\cite{PhysRevC.21.2293,PhysRevC.19.1276,PhysRevC.18.86}.
\par
\begin{figure}[t]
	\centering
  \includegraphics[width=0.45\textwidth]{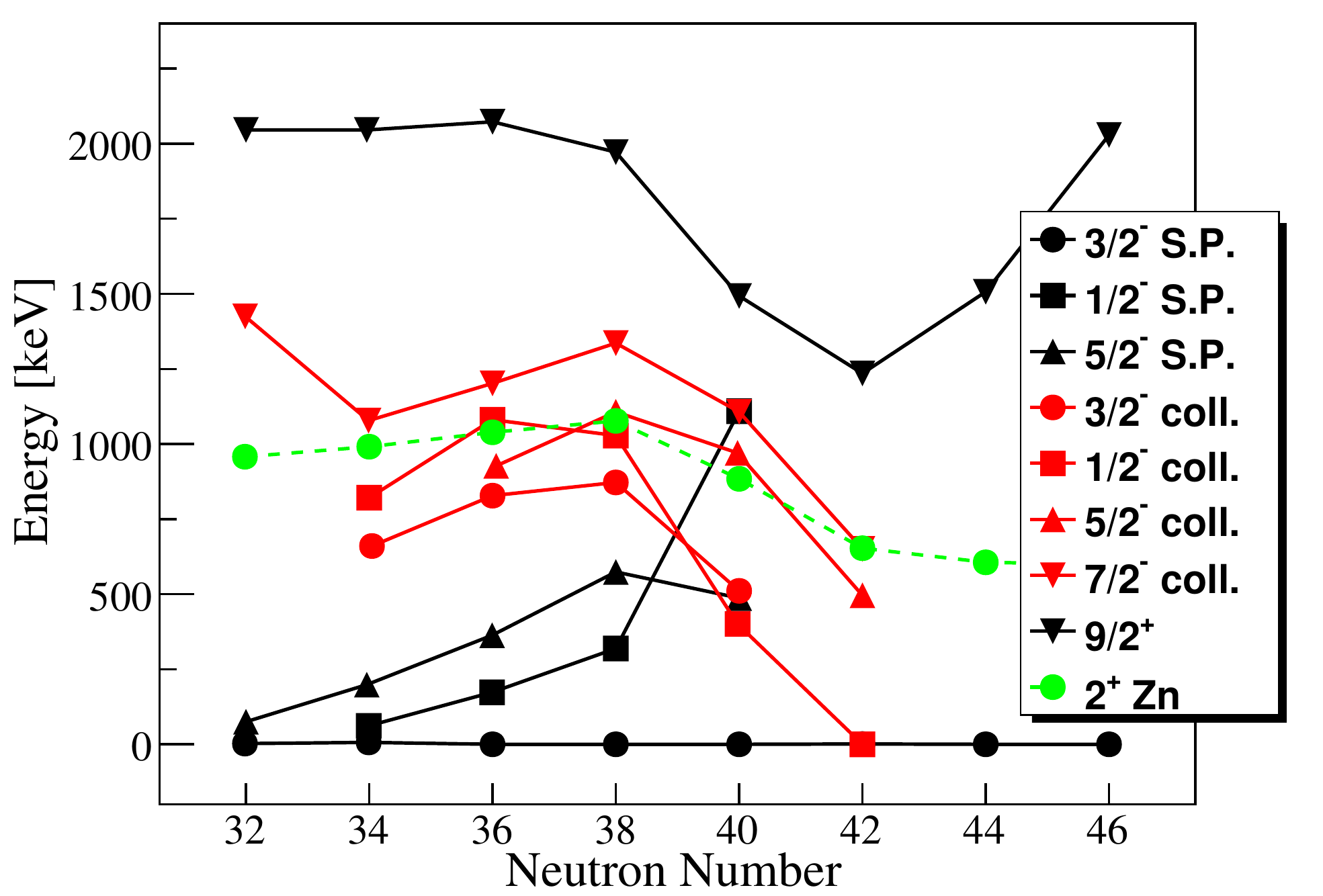}
  \caption{(Color) Overview of the odd-A gallium isotopes level energy systematics. The black data points indicate with a single particle character, while the red points indicate collective states. The excitation energy of the first $2^+$ state in the zinc isotopes is shown in green. Data taken from~\cite{nndc}}
  \label{fig:systematics}
\end{figure}

From the energy systematics, there seems to be a more general structural difference between the odd-A gallium isotopes up to ${}^{69}$Ga${}_{38}$ and those with 40 neutrons or more (see figure~\ref{fig:systematics}). The level structure at low excitation energy of the former is characterized by 2 groups of levels: one consists of the 3/2${}^-$ ground state along with a 1/2${}^-$ and 5/2${}^-$ level at low excitation energy, which are all strongly populated in one proton transfer reactions on zinc isotopes, suggesting a $\pi pf$ single-particle character~\cite{Riccato1974461,PhysRevC.9.409}. The second group of levels comprises a 1/2${}^-$,3/2${}^-$,5/2${}^-$ and 7/2${}^-$ multiplet, situated around an energy of 1 MeV and can originate mainly from the $\pi p_{3/2}\otimes$2${}^+$(Zn) core-coupling scheme. The average excitation energy of the members of this multiplet agrees very well with the energy of the $2^+_1$ state in the corresponding zinc isotopes. Measured $B(E2,\downarrow)$-values for ground state transitions from members of this multiplet in these isotopes are similar in magnitude to B(E2;$2^+_1\rightarrow 0^+$)-values in the corresponding zinc core supporting this assignment as the dominant component of the wave function~\cite{Stelson1962652,andreev}. 
\par
In ${}^{71}$Ga${}_{40}$ the separation between these two groups of levels is less obvious. Also note from figure~\ref{fig:systematics} that the $9/2^+_1$ level reaches a minimum in excitation energy for $N=42$. Calculations within the Coriolis coupling model including the pairing interaction reproduce this minimum~\cite{PhysRev.176.1355}. A possible explanation for this trend is that increased deformation in the $N=28-50$ mid-shell region brings the $\pi 1g_{9/2}$ orbital down in energy~\cite{irina_deep_inelastic}.
\par 
In order to get a better understanding of the structural changes in the gallium isotopes between $N = 40$ and $42$ and to find evidence for the proposed ground state doublet, a Coulomb excitation experiment was performed using post-accelerated beams of ${}^{71,73}$Ga~\cite{alder}. 

\section{EXPERIMENTAL SET-UP}
The radioactive ${}^{71,73}$Ga ion beam were produced at the REX-ISOLDE facility in CERN by irradiating a 45 g/cm${}^{2}$ UC${}_x$ target with a 1.4 GeV pulsed proton beam with an intensity of $3\times 10^{13}$ protons/pulse (18 pulses/minute). The primary target matrix was kept at a temperature of approximately 2000 ${}^{\circ}$C to optimize the release time of the ions from the source. The gallium atoms were surface ionized in a hot cavity and were extracted from the source by applying a 30 kV extraction potential. Beside surface ionization of gallium atoms also copper atoms were selectively ionized by resonant laser ionization as the Coulomb excitation of neutron-rich copper isotopes was one of the topics of interest~\cite{rilis}. After mass separation the ions of interest were bunched by REX-TRAP, charge bred in REX-EBIS to a charge state of ${19}^+$, post-accelerated to an energy of 2.95 MeV/nucleon by the REX linear accelerator~\cite{voulot} and finally directed on a 1.7 mg/cm${}^2$ ${}^{120}$Sn or a 2 mg/cm${}^2$ ${}^{104}$Pd target to induce Coulomb excitation.\par
De-exciting $\gamma$ rays after Coulomb excitation were detected by the MINIBALL $\gamma$-detector array which consists of eight clusters of three HPGe crystals which are all electrically sixfold segmented~\cite{eberth}. The absolute photo peak efficiency of this detector array is 7.2 \% at 1332 keV. A 500 $\mu$m thick CD-shaped segmented double-sided silicon detector was placed 35 mm behind the target to register the scattered projectiles and recoiling target nuclei~\cite{ostrowski}. This detector is divided in four quadrants which have 16 angular rings in the front and 24 sector strips in the back, covering an angular range in the laboratory system between 16${}^{\circ}$ and 53${}^{\circ}$.\par
The beam composition was monitored by comparing laser on (Ga + Cu) with laser off (only Ga) measurements on a regular basis. With lasers on, the average ${}^{73}$Ga-to-total ratio of the beam during the experiment was 82.6(6) \%, with a ${}^{73}$Ga intensity at the MINIBALL set-up of $3.8\times 10^5$ pps. For mass 71 the average ${}^{71}$Ga-to-total ratio was equal to 35(1) \% with a ${}^{71}$Ga intensity of $1.2\times 10^5$ pps.
\begin{table}[tb]
	\centering
		\begin{tabular}{cccccc}
		\hline
		\hline
		\multirow{2}{*}{Mass} & \multirow{2}{*}{$E_{\gamma}$ [keV]} & \multirow{2}{*}{$I^{\pi}_{i} \rightarrow I^{\pi}_{f}$} & \multicolumn{3}{c}{$B(E2,\downarrow)$ [W.u.]} \\		
		& &  & This work & ~\cite{andreev} & Avg \\
		\hline
		
		\multirow{6}{*}{${}^{69}$Ga} 	& 319 & $1/2^-_1 \rightarrow 3/2^-_1$ &      & 4.0(2) & \\
		& 574 & $5/2^-_1 \rightarrow 3/2^-_1$ &      & 0.28(5) & \\
		& 872 & $3/2^-_2 \rightarrow 3/2^-_1$ &      & 4(1) & \\
		& 1029 & $(1/2^-_2) \rightarrow 3/2^-_1$ &      & 3.4(9) & \\
		& 1107 & $5/2^-_2 \rightarrow 3/2^-_1$ &      & 18(2) & \\
		& 1337 & $7/2^-_1 \rightarrow 3/2^-_1$ &      & 10(1) & \\
		
		\hline
		
	\multirow{7}{*}{${}^{71}$Ga} &	390 & $1/2^-_1 \rightarrow 3/2^-_1$ &      & $<$ 0.97 & \\
		& 487 & $5/2^-_1 \rightarrow 3/2^-_1$ &      & $<$ 0.12 & \\ 
		& 512 & $3/2^-_2 \rightarrow 3/2^-_1$ & 5(2) & 4.6(7) & 4.6(6) \\
		& 965 & $5/2^-_2 \rightarrow 3/2^-_1$ & 9(5) & 13(4) & 11(3)\\		
		& 1107 & $7/2^-_1 \rightarrow 3/2^-_1$ &     & 0.8(1) & \\
		& 1109 & $1/2^-_2 \rightarrow 3/2^-_1$ &     & 7.2(12) & \\
		& 1395 & $5/2^-_3 \rightarrow 3/2^-_1$ &     & 3.7(5) & \\
		
		\hline
		\hline
		\end{tabular}
\caption{Overview of information on $B(E2)$-values in the neutron-rich gallium isotopes, including the extracted $B(E2)$-values for ${}^{71}$Ga from this work and those reported in~\cite{andreev}. The large uncertainty on the presented $B(E2)$-values is due to the limited amount of statistics (93 $\pm$ 25 counts (512 keV) and 67 $\pm$ 26 counts (965 keV)).}
\label{tab:71ga}
\end{table}
\section{ANALYSIS}
The analysis of the Coulomb excitation data on ${}^{69,71,73}$Cu was published previously in~\cite{stefanescu}. This work show the population of $1/2^-$, $5/2^-$ and $7/2^-$ levels. $B(E2)$-values indicate that three different types of states exist at low excitation energy in the neutron-rich copper isotopes above $N = 40$ in contrast to mainly core-coupled structures below $N = 40$~\cite{robinson}. These three modes are a core-coupled $7/2^-$ state, a $5/2^-$ state with a strong single-particle character and a collective $1/2^-$ state.
\par

\begin{figure}[tb]
 \centering
 \includegraphics[width=0.45\textwidth]{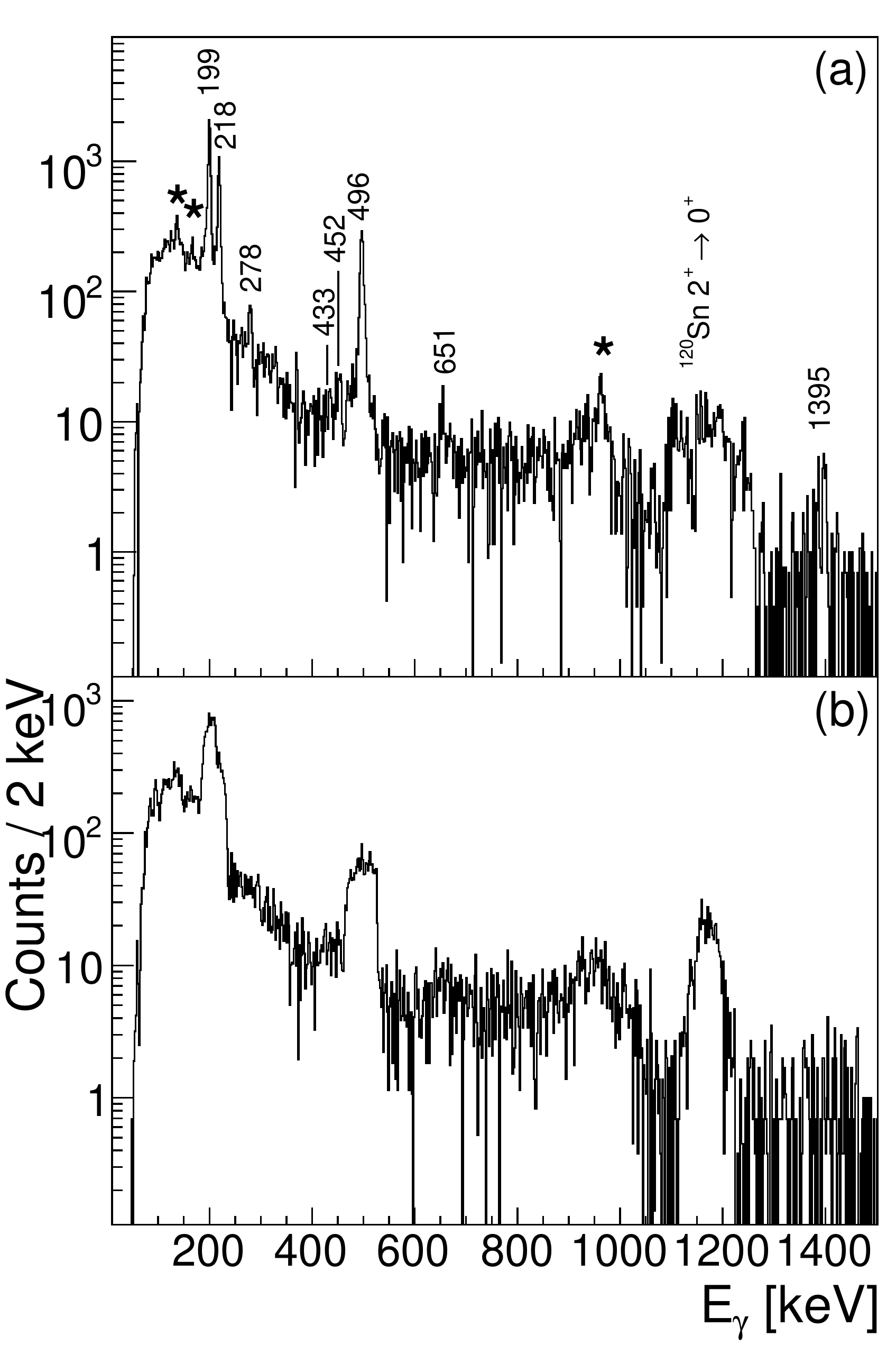}
 \caption{Particle coincident $\gamma$-ray spectrum Doppler corrected (a) and non Doppler corrected (b) for projectile ($A=73$) excitation. Transitions with an energy indication (in keV) originate from depopulation of levels in ${}^{73}$Ga and these transitions are tabulated in table~\ref{tab:73gammas}, while the transitions marked with an asterisk are transitions in ${}^{73}$Cu.}
 \label{fig:spectrumDC73}
 \end{figure}

\begin{table*}[tb]
	\centering
		\begin{tabular}{ccccccccccc}
		\hline
		\hline
		 & & & & \multicolumn{5}{c}{$E_{\gamma}$ Previously observed [keV]} & ISOLDE & Adopted\\
		E${}_{\gamma}$ [keV] & $I^{\pi}_{i}$ & $I^{\pi}_{f}$ & Counts & \cite{Erdal1972449} & \cite{Runte1983163} & \cite{PhysRevC.19.1276} & \cite{PhysRevC.21.2293} & \cite{irina_deep_inelastic} & $\beta$-decay & Value \\
		\hline
		199.2(5)  & $5/2^-_1$ & ${3/2^-_1}^{(*)}$ & 5530 (102)   & 				& 					& 			 	& 198(3)	& 199.1(2) 		&  					& 199.1(2) \\
		218.4(6)  & $3/2^-_2$ & ${1/2^-_1}^{(*)}$ & 2294 (66)    & 216(2) & 218.1(2)	& 219(3) 	& 216(3) 	& 218.2(2)		& 217.9(2) 	& 218.08(11) \\
		279.0(7)  & $5/2^-_2$ & $3/2^-_2$ 				& 161 (29)     & 				& 278.4(4)	& 				& 				&  						&  					& 278.5(3)\\
		298(2)    & $5/2^-_2$ & $5/2^-_1$         & -            &        &						&					&					&							&						& 298 (2) \\  
		434.0(15) & $7/2^-_1$ & $3/2^-_2$ 				& 23 (14)      & 				&						&	 				& 				& 433.0(5) 		& 					& 433.1(5)\\
		451.7(11) & $7/2^-_1$ & $5/2^-_1$ 				& 68 (19)    	 &				&						& 				& 				& 452.1(2) 		&  					& 452.1(2)\\
		495.8(5)  & $5/2^-_2$ & ${3/2^-_1}^{(*)}$ & 1187 (34)    & 496(2) & 495.6(3) 	& 498(3)	& 495   	& 496.2(2) 		& 496.2(2) 	& 496.07(12)\\
		651(2)    & $7/2^-_1$ & $3/2^-_1$ 				& 49 (9) 			 & 				&						& 				& 				& 651.2(2) 		&  					& 651.2(2)\\
		          & $3/2^-_3$ & $3/2^-_2$ 				&              &        & 693.1(3) 	&         &         &             & 693.3(3) 	& 693.2(2) \\
		          & $3/2^-_3$ & $3/2^-_1,1/2^-_1$ &      				 & 911(3) & 910.5(4) 	& 915(3) 	&   913(4)&             & 911.4(2) 	& 911.2(2) \\	
		1395.1(12) & $5/2^-_3$ & $3/2^-_1,1/2^-_1$& 42 (7) 		 	 & 				&						& 1396(3)	&  				& 						&  					& 1395.2(11)\\  
		\hline
		\hline
		\end{tabular}
\caption{Overview of the selected $\gamma$ transitions in ${}^{73}$Ga. Previously known $\gamma$-ray energies and available uncertainties are taken from~\cite{Erdal1972449,Runte1983163,PhysRevC.21.2293,PhysRevC.19.1276,irina_deep_inelastic}. Number of counts are extracted from fig.~\ref{fig:spectrumDC73}. ${}^{(*)}$Proposed $I^{\pi}_{f}$ based on a comparison with similar transitions observed in ${}^{75}$Ga.}
\label{tab:73gammas}
\end{table*}
The gallium data were analyzed in a similar way as the copper data by requiring particle-$\gamma$ coincidences. By using the position information from both the particle and $\gamma$ detection a Doppler correction could be performed. In this way two transitions with energies of 199 keV and 218 keV (figure~\ref{fig:spectrumDC73}.a) could be resolved from the broad structure around $200$ keV in the non Doppler corrected spectrum (figure~\ref{fig:spectrumDC73}.b).\par
The available data on ${}^{71}$Ga were analyzed to serve as a proof of principle and validate the procedure as the extracted $B(E2)$-values can be compared to the values published by Andreev \textit{et al.}~\cite{andreev}. Only data from laser on runs were used as the amount of laser off data, giving a pure gallium beam, was very limited. The number of observed ${}^{104}$Pd target excitations had to be corrected for excitations caused by copper. The procedure which was applied is described extensively in~\cite{walle:014309}. By using the GOSIA2 analysis code~\cite{gosia} the relevant matrix elements could be extracted and both the results from this analysis and values from literature~\cite{andreev} are presented in table~\ref{tab:71ga} along with their weighted averages. Despite the limited statistics the obtained results agree with those published in~\cite{andreev}. The error analysis includes both statistical uncertainties as well as systematical contributions arising from uncertainties in the target $B(E2)$-values and $\gamma$-detection efficiency.\par
\par
The analysis of the ${}^{73}$Ga data was very similar to the one outlined before, with the difference that a ${}^{120}$Sn target was used for normalization. Also only laser on data were used to maximize the statistics. From the measured Doppler corrected particle coincident $\gamma$-ray spectrum (figure~\ref{fig:spectrumDC73}.a) energies and intensities could be extracted and are summarized in table~\ref{tab:73gammas}, together with values from literature and weighted averages.\par
While in the Coulomb excitation process, the population of the excited states is by far dominated by the $E2$ matrix elements connecting the ground state with the excited states, the decay is governed by the possible decay paths expressed in $\gamma$-decay branching ratios and possible mixing ratios, here mainly of $M1/E2$ character. This additional degree of freedom has as a result that the $E2$ matrix elements between excited states in ${}^{73}$Ga cannot be firmly fixed. The influence of these matrix elements on the extracted $B(E2)$-values was extensively verified and included in the uncertainty on the values presented in table~\ref{tab:BE2_73Ga}. In the analysis the quadrupole moments of all states were assumed to be $0$ eb. 

 \begin{figure}[tb]
	\centering
		\includegraphics[width=0.45\textwidth]{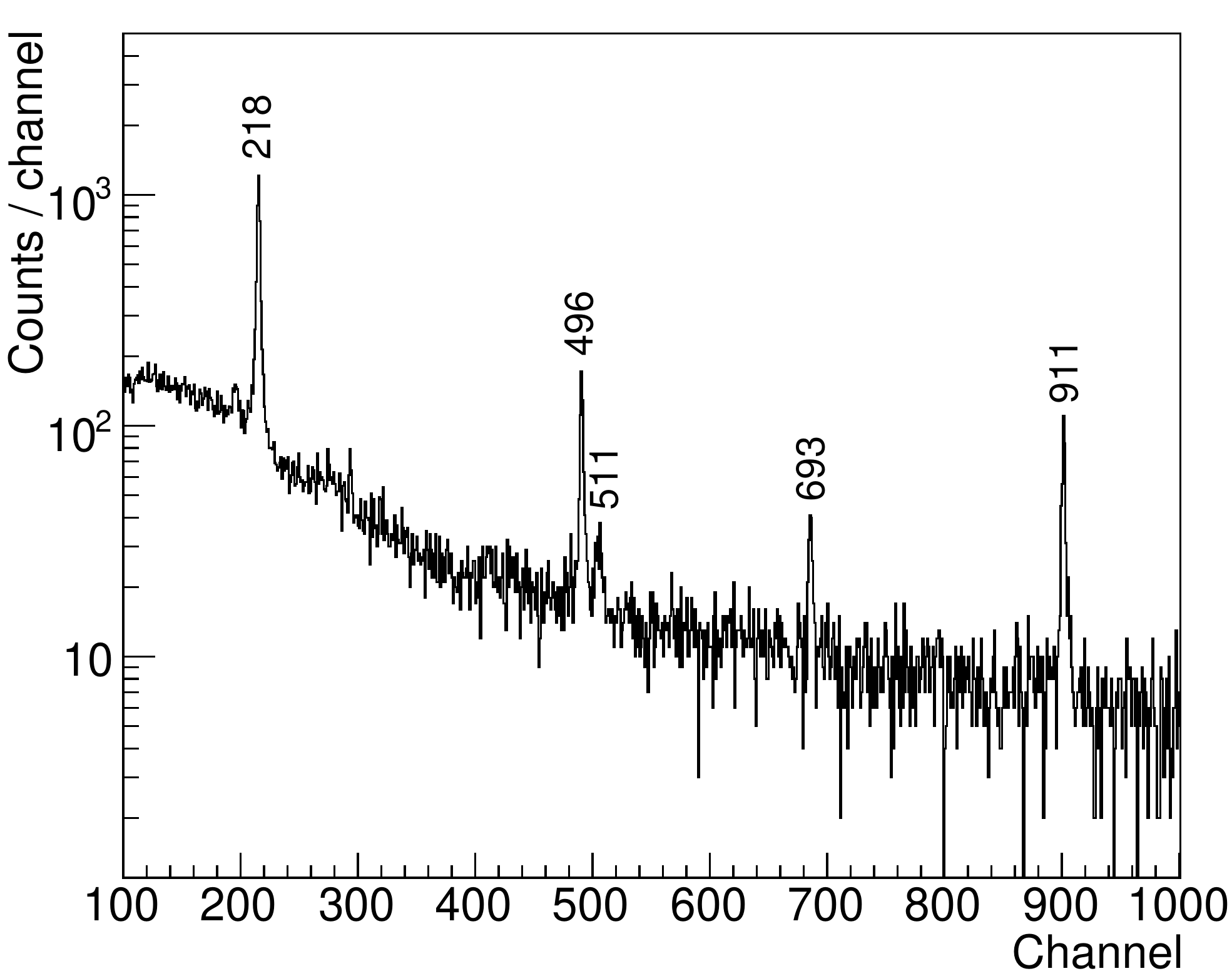}
	\caption{${}^{73}$Zn $\beta$-decay spectrum obtained using the ISOLDE tape station. Energy given in keV.}
	\label{fig:prc_beta}
\end{figure}

\begin{table}[tb]
	\centering
		\begin{tabular}{cccc}
		\hline
		\hline
		$E_{\textrm{state}}$ [keV] & $I^{\pi}_{i}$ & $I^{\pi}_{f}$ & $B(E2,\downarrow)$ [W.u.] \\		
		\hline
		199${}^{*}$ & $5/2^-_1$ & $1/2^-_1$  & 11(2) \\
		218 				& $3/2^-_2$ & $1/2^-_1$ & 7.5 (10) \\
		496${}^{*}$ & $5/2^-_2$ & $1/2^-_1$ & 6.5 (10) \\
		1395 				& $5/2^-_3$ & $1/2^-_1$ & 3.0(7) \\
		\hline
		\hline
		\end{tabular}
\caption{Overview of the extracted $B(E2)$-values for ${}^{73}$Ga. ${}^{*}$ indicates states where the $\gamma$ decay populates the $3/2^-$ member of the ground state doublet. Hence the excitation energy of these states is equal to the energy of the observed depopulating $\gamma$ ray plus the $3/2^-$-$1/2^-$ energy difference that is currently only known as an upper limit.}
\label{tab:BE2_73Ga}
\end{table}
\section{RESULTS AND DISCUSSION}

\begin{figure}
\centering
\includegraphics[width=0.45\textwidth]{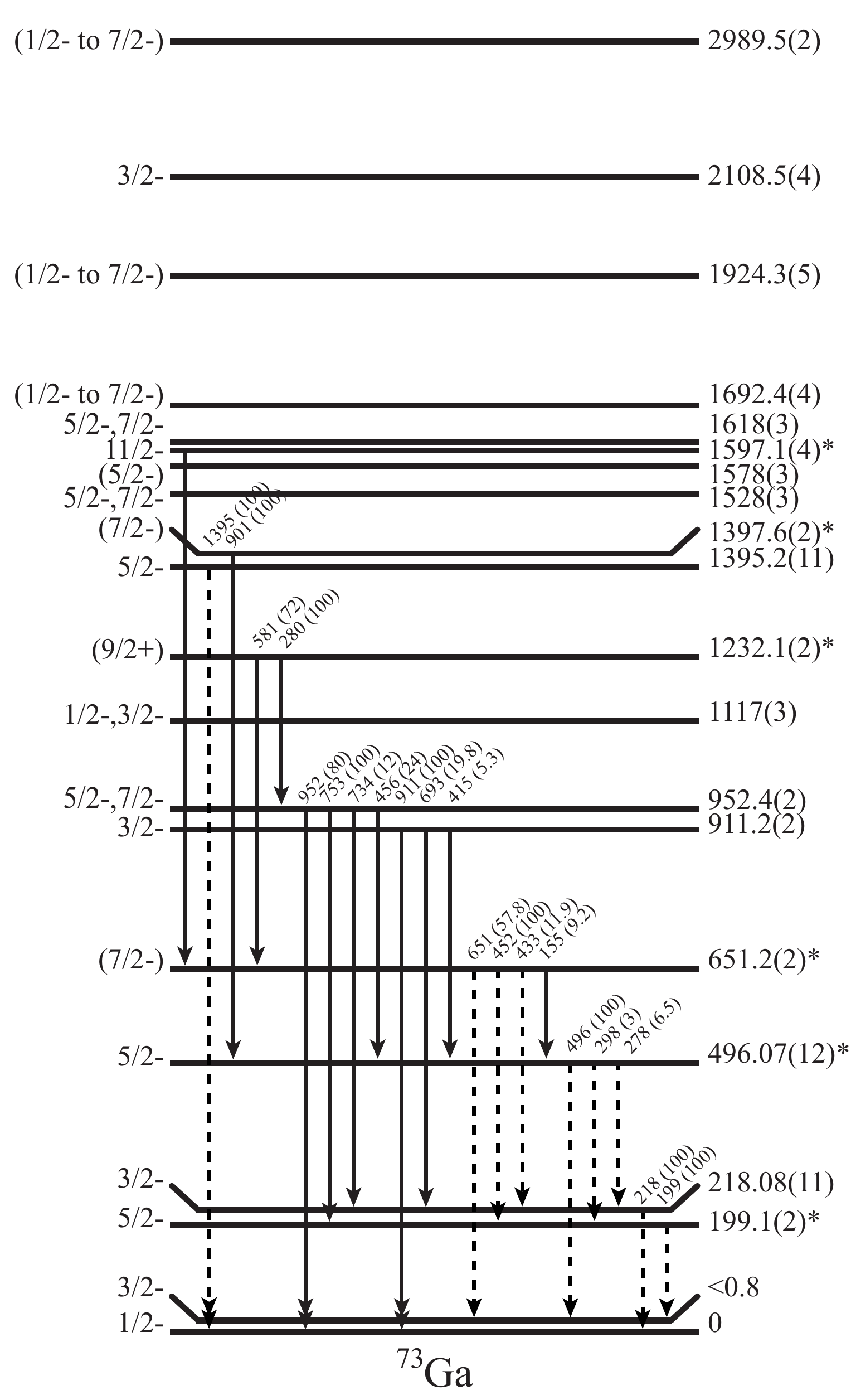}
\caption{Level and $\gamma$-decay scheme of ${}^{73}$Ga. Transitions observed during the Coulomb excitation experiment are depicted as a dashed line. The energy of the levels indicated with a * should be increased by the $3/2^-$-$1/2^-$ energy difference that is currently only known as an upper limit. Data taken from ENSDF~\cite{nndc} and this work.}
\label{fig:73ga_scheme}
\end{figure}

In figure~\ref{fig:73ga_scheme} the level structure of ${}^{73}$Ga is shown based on literature~\cite{nndc} and this work. Considering recent laser spectroscopy data the ground state spin was found to be $1/2^-$ while transfer reactions gave evidence for a $3/2^-$ state close to the ground state. No evidence for a doublet near the ground state of ${}^{73}$Ga was obtained from the transfer reaction studies~\cite{PhysRevC.18.86,Riccato1974461,PhysRevC.19.1276,PhysRevC.21.2293,PhysRevC.2.149,PhysRevC.9.409}.
\par
The results obtained from the analysis of the ${}^{73}$Ga Coulomb excitation data support the existence of this doublet. From figure~\ref{fig:spectrumDC73}.b it can be seen that the peak originating from the depopulation of the 199 keV state is Doppler broadened. An excited state with an energy of 198(3) keV was observed previously in the one proton stripping reaction with $\Delta$L = 3, restricting its spin and parity to 5/2${}^-$,7/2${}^-$~\cite{PhysRevC.21.2293}. As this state is populated directly by Coulomb excitation the spin can be firmly fixed as 5/2${}^-$. A recent ${}^{74}$Ge(d,${}^{3}$He) transfer reaction using polarized deuterons also favors a 5/2${}^{-}$ assignment for this state~\cite{Kay,PhysRevC.79.021301}.\par
The observed Doppler broadening imposes a restriction on the lifetime of this 5/2${}^-_1$ state. As the scattered particles are all stopped in the CD detector, the lifetime of this level should be considerably smaller than the 3.5 ns maximum time of flight between the ${}^{120}$Sn target and the CD detector. The lifetime of this 5/2${}^-_{1}$ state in case of a pure E2 decay to the 1/2${}^-$ ground state can be calculated to be 13(2) ns based on the experimental $B(E2;5/2^-_1 \rightarrow 1/2^-_1)$-value found in table~\ref{tab:BE2_73Ga}. The observed Doppler broadening of the 199 keV peak can thus only be explained if it has an important, fast M1 component. Note that the Weisskopf estimate for the half-life of this 5/2${}^-_{1}$ state for a pure 199 keV M1 transition is 2.3 ps. We can thus conclude that there exists a state very close to the $1/2^-$ ground state in ${}^{73}$Ga with spin and parity ranging between $3/2^-$ and $7/2^-$. The two-neutron transfer reaction indicates a $L=0$ character~\cite{PhysRevC.19.1276} thus the low lying state that forms a doublet with the $1/2^-$ state must have spin and parity $3/2^-$. The inverse process has already been applied in a similar experiment to show that specific states have lifetimes in the range of nanoseconds, which was used to deduce the multipole character of the de-exciting $\gamma$ ray.~\cite{stefanescu_oddodd}.
\par
Also gamma rays de-exciting levels at 218 keV and 496 keV were observed and the deduced Coulomb excitation cross sections are compatible with single-step Coulomb excitation. From two-neutron transfer reaction data, the spin and parity of the 218 keV level are known to be 3/2${}^{-}$. A state around 495 keV was observed in the ${}^{74}$Ge(d,${}^{3}$He) reaction with $\Delta$L = 3~\cite{PhysRevC.21.2293}, limiting the spin and parity of this state to 5/2${}^-$ or 7/2${}^-$. Data from the ${}^{74}$Ge(d,${}^{3}$He) reaction using polarized deuterons favor a 7/2${}^-$ assignment for this state.~\cite{Kay}. The fact that we observe direct population of the 496 keV state via Coulomb excitation from a 1/2${}^{-}$ ground state however limits the spin of this state to be 5/2${}^{-}$. Also the lack of $\gamma$ feeding towards this state in the decay of the 9/2${}^{+}_{1}$ state, in contrast with the two 7/2${}^{-}$-states at 952 keV and 651 keV~\cite{irina_deep_inelastic}, supports this 5/2${}^-$ assignment. However, the possibility of a 5/2${}^-$,7/2${}^-$ doublet around 495 keV cannot be completely excluded.\par
The energy difference between the two members of the ground state doublet was estimated by comparing $\gamma$-ray energies of different decay paths of excited states that preferentially populate a different member of the doublet. A possible candidate is the decay of the 496 keV 5/2${}^-_2$ level assuming the $\gamma$ decay branchings in ${}^{73}$Ga are similar to those in ${}^{75}$Ga. From the intensity ratios $I(5/2^-_2\rightarrow 1/2^-_1):I(5/2^-_2\rightarrow 3/2^-_1) = 0.36$ and $I(3/2^-_2\rightarrow 1/2^-_1):I(3/2^-_2\rightarrow 3/2^-_1) = 29$ in ${}^{75}$Ga it can be expected that the 496 keV transition in ${}^{73}$Ga will preferentially populate the 3/2${}^-$ member of the doublet by a M1 transition, while the cascade of 278 keV and 218 keV transitions primarily feeds the 1/2${}^-$ state~\cite{Ekstrom}. In order to improve the precision of the $\gamma$-ray energies used, available ${}^{73}$Zn $\beta$-decay data from the ISOLDE tape station~\cite{walle:014309} were analyzed (see figure~\ref{fig:prc_beta}) and included in the adopted values tabulated in table~\ref{tab:73gammas}. The adopted $\gamma$-ray energy of the direct decay is 496.07(12) keV (see table~\ref{tab:73gammas}), while the indirect branch consists of two $\gamma$ rays with energies of 218.08(11) keV and 278.5(3) keV adding up to 496.5(4) keV. This results in an energy difference of 0.4(4) keV between the two decay paths. This imposes an upper limit on the excitation energy of the $3/2^-_1$ state of 0.8 keV within 1 $\sigma$. 
\par
Detailed analysis of the 218 keV peak shape (figure~\ref{fig:prc_beta}) does not show any broadening of the peak, indicating that the transition is either mono-energetic or that the energy difference is too small to cause any peak asymmetry.
\par
Also the population of a level with an energy of 1395.1(12) keV in ${}^{73}$Ga is observed. This level was observed before in the ${}^{71}$Ga(t,p)${}^{73}$Ga reaction, with an energy of 1395(3) keV and $\Delta L=4$ angular momentum transfer, yielding spin and parity possibilities ranging from 5/2${}^-$ up to 11/2${}^-$~\cite{PhysRevC.19.1276}. The observed population of this 1395 keV level is compatible with single step Coulomb excitation as a two step process would require unphysically large $B(E2)$-values between excited $3/2^-,5/2^-$ states and this 1395 keV state. In the case of two step Coulomb excitation through the $5/2^-_2$ level, the required $B(E2)$-values would be $B(E2;5/2^-_2 \rightarrow 5/2^-_3)$ = 17000 e${}^2$fm${}^4$ or $B(E2;5/2^-_2 \rightarrow 9/2^-_1)$ = 8000 e${}^2$fm${}^4$ depending on the spin of the 1395 keV state. Such $B(E2)$-values would lead to strong $\gamma$-decay branches to excited $3/2^-, 5/2^-$ states in comparison with the branch to the doublet near $0$ keV excitation energy. Hence, the lack of strong $\gamma$ branches to the excited $3/2^-$ and $5/2^-$ states also supports the interpretation in terms of single step Coulomb excitation, constraining the spin and parity of this level to $5/2^-$.
\par
In addition, a series of weaker transitions with energies of 433, 452 and 651 keV is observed. These originate from the de-excitation of a level at 651.2 keV which was recently proposed with a spin and parity assignment of 7/2${}^-$~\cite{irina_deep_inelastic}. As M3 excitations are strongly hindered during the Coulomb excitation process, the population of this state by Coulomb excitation from a 1/2${}^-$ ground state can only take place by a two step process. While the population of this level is governed by $\left<3/2^-,5/2^-\left|E2\right|7/2^-_1\right>$ matrix elements, the decay is determined by both these $E2$- and the $\left<5/2^-\left|M1\right|7/2^-_1\right>$ matrix element. Due to the high number of degrees of freedom, $E2$ matrix elements between $3/2^-,5/2^-$ states and this $7/2^-_1$ state cannot be firmly fixed in our analysis. Hence, $B(E2;7/2^-_1 \rightarrow 3/2^-,5/2^-)$-values for the observed transitions are not included in table~\ref{tab:BE2_73Ga}, while the influence of these transitions on the published $B(E2)$-values in table~\ref{tab:BE2_73Ga} was extensively investigated and included in the error bars. 
\par
Also the possible influence of multistep Coulomb excitation via the $3/2^-_1$ level on the extracted $B(E2)$-values was verified and the effect was incorporated in the uncertainties presented in table~\ref{tab:BE2_73Ga}.
\par
\begin{figure}[tb]
\centering
 \includegraphics[width=0.45\textwidth]{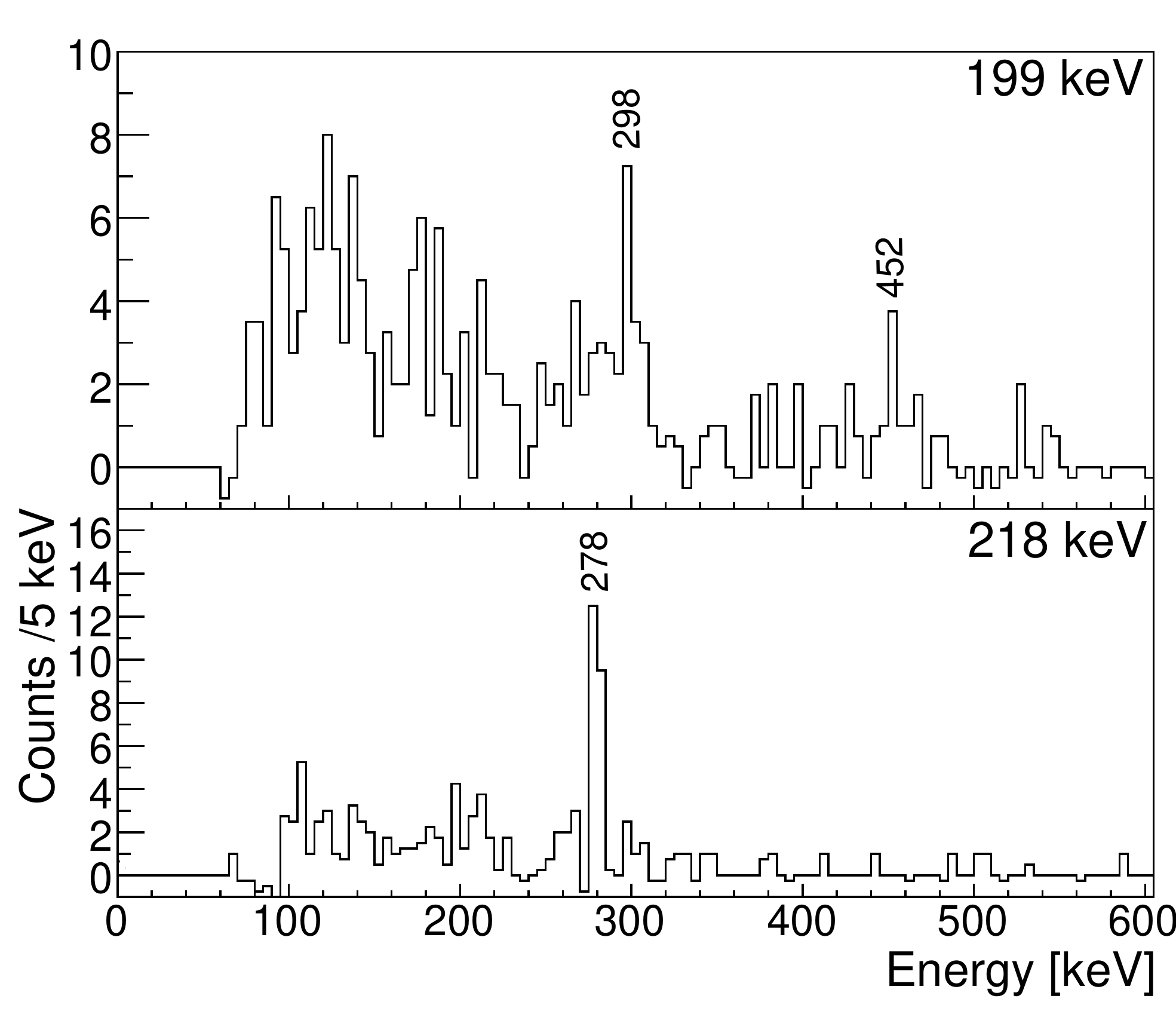}
 \caption{Doppler corrected $\gamma$-spectra coincident with the 199 keV (top) and the 218 keV (bottom) transition. The 199 keV coincidence spectrum shows coincidences with both 298 keV and 452 keV transitions, while the 218 keV $\gamma$ ray is coincident with 278 keV. No coincidence between the 218 keV and 433 keV $\gamma$ rays is observed, which is due to the relatively small 433 keV $\gamma$ branch in the decay of the 651 keV state~\cite{irina_deep_inelastic}. Based on the limited number of 452-199 keV coincidences observed the top spectrum less then one 433-218 keV coincidence count is expected in the lower spectrum.}

 \label{fig:coinc73}
 \end{figure}
Analysis of particle-$\gamma$-$\gamma$ events shows a new branch in the decay of the 496 keV 5/2${}^-_2$ level. The coincidence spectrum (see Fig.~\ref{fig:coinc73}) shows beside the known coincidence of the 218 keV and 278 keV transitions also a coincidence between the 199 keV and 298 keV transitions. The new transition with an energy of 298(2) keV can be inserted into the level scheme as the transition between the 496 keV 5/2${}^-_2$ and 199 keV 5/2${}^-_1$ levels with a branching ratio of 3(1)\% relative to the 496 keV transition.
\par
The extracted $B(E2,\downarrow)$-values support a shape transition between ${}^{71}$Ga${}_{40}$ and ${}^{73}$Ga${}_{42}$ as a shift of the collectivity towards lower excitation energy is observed. A measure for the quadrupole collectivity can be obtained by calculating the $B(E2)$ weighted energy of the observed excitations ($E_{centroid} = \sum {B(E2).E} /\sum{B(E2)}$). Values of 1039 keV (${}^{67}$Ga${}_{36}$), 1062 keV (${}^{69}$Ga${}_{38}$), 956 keV (${}^{71}$Ga${}_{40}$) and 402 keV (${}^{73}$Ga${}_{42}$) were obtained, clearly indicating an abrupt lowering of the average energy of the E2 strength beyond $N=40$. When compared to the energy of the first $2^{+}$ state in the zinc isotopes (1039 keV (${}^{66}$Zn${}_{36}$), 1077 keV (${}^{68}$Zn${}_{38}$), 884 keV (${}^{70}$Zn${}_{40}$) and 653 keV (${}^{72}$Zn${}_{42}$)) it can be seen that the energy of the E2 strength clearly deviates from the zinc $E(2^{+}_{1})$ at $N=42$.\par
Systematic studies in the neutron rich copper isotopes have shown a drop in excitation energy of the yrast $1/2^-$ level beyond $N=40$, while the increasing $B(E2;1/2^-_1 \rightarrow 3/2^-_1)$ indicates enhanced deformation~\cite{stefanescu}. The similarity between the measured $B(E2; 1/2^-_1 \rightarrow 3/2^-_2)$-value of $15(2)$ W.u. in ${}^{73}$Ga${}_{42}$ and the $B(E2; 1/2^-_1 \rightarrow 3/2^-_1)$-value of $20.4(22)$ W.u. in ${}^{71}$Cu${}_{42}$ might indicate the resemblance of the structure of this $1/2^-$ level in both nuclei and indeed point to enhanced ground state deformation in ${}^{73}$Ga.
\section{SUMMARY}
A Coulomb excitation experiment was performed using post accelerated beams of ${}^{71,73}$Ga. Information on the ${}^{73}$Ga level scheme has been extended. New evidence for the existence of a 1/2${}^-$, 3/2${}^-$ ground state doublet could be extracted based on the lifetime of the 199 keV 5/2${}^-$ level. An upper limit of $0.8$ keV can be imposed on the excitation energy of the $3/2^-_1$ state. The abrupt lowering of the energy of the quadrupole collectivity was observed and support a shape change between ${}^{71}$Ga${}_{40}$ and ${}^{73}$Ga${}_{42}$, highlighting the importance of the neutron $g_{9/2}$ orbital in inducing collectivity in this region. Extending these measurements towards more neutron rich gallium isotopes would allow one to investigate the stabilizing effect of the $N=50$ shell closure. 

\begin{acknowledgments}
This work was supported by the European Commission within the Sixth Framework Programme through I3-EURONS (Contract RII3-CT-2004-506065), FWO-Vlaanderen (Belgium), GOA/2004/03 (BOF-K.U.Leuven), the 'Interuniversity Attraction Poles Programme - Belgian State - Belgian Science Policy'  (BriX network P6/23), HIC for FAIR and the BMBF under grants 06KY9136I, 06MT238, 06MT9156 and 06DA9036I.
\end{acknowledgments}


\end{document}